\documentclass
[superscriptaddress,secnumarabic,amssymb,amsmath,nobibnotes,aps,prd,showkeys,showpacs,nofootinbib]{revtex4}%
\usepackage{graphicx}
\usepackage{epsf}
\usepackage{bm}
\usepackage{amsmath}
\usepackage{amsfonts}
\usepackage{amssymb}
\usepackage{epstopdf}%
\providecommand{\U}[1]{\protect\rule{.1in}{.1in}}
\newcommand{\be}{\begin{equation}}
\newcommand{\ee}{\end{equation}}

\newcommand{\mincir}{\raise
-3.truept\hbox{\rlap{\hbox{$\sim$}}\raise4.truept\hbox{$<$}\ }}
\newcommand{\magcir}{\raise
-3.truept\hbox{\rlap{\hbox{$\sim$}}\raise4.truept\hbox{$>$}\ }}

\begin{document}
\title{Similarity solutions for the Wheeler-DeWitt equation in $f\left(  R\right)  $-cosmology}
\author{Andronikos Paliathanasis}
\email{anpaliat@phys.uoa.gr}
\affiliation{Institute of Systems Science, Durban University of Technology, PO Box 1334,
Durban 4000, RSA}


\begin{abstract}
In the case of a spatially flat Friedmann--Lema\^{\i}tre--Robertson--Walker
Universe in $f\left(  R\right)  $-gravity we write the Wheeler-DeWitt equation
of quantum cosmology. The equation depends upon the functional form of
$f\left(  R\right)  $. We choose to work with four specific functions of
$f\left(  R\right)  $ in which the field equations for the classical models
are integrable and solvable through quadratures. For these models we determine
similarity solutions for the Wheeler-DeWitt equation by determining
Lie-B\"{a}cklund transformations. In addition we show how the classical limit
is recovered by the similarity solutions of the Wheeler-DeWitt equation.

\end{abstract}
\maketitle

\section{Introduction}

Modified theories of gravity \cite{cl1,cl2} are an alternate approach to the
dark energy models to explain recent observational phenomena
\cite{Teg,Kowal,Komatsu,planck}. The common characteristic of the modified
theories of gravity is the modification of the Einstein-Hilbert Action by
adding new invariant terms to the gravitational Action. The novelty of that
approach is that new geometrodynamical components are introduced into the
field equations which drive the dynamics to explain the observations.

In the literature there have been proposed a plethora of different modified
theories of gravity. A specific class of models which have drawn the attention
are the so-called $f-$theories. In $f-$theories of gravity a function
$f\left(  Q\right)  $ is introduced into the Einstein-Hilbert Action, where
$Q$ is an invariant function. Some theories which belong to that class of
models are the: $f\left(  R\right)  $-gravity in the metric formalism
\cite{fr1}, $f\left(  \hat{R}\right)  $-gravity in the Palatini formalism
\cite{pal}, $f\left(  G\right)  $-Gauss Bonnet theory \cite{bli}, while in the
Teleparallel formalism of gravity the $f\left(  T\right)  -$theory has been
widely studied the last decade \cite{ft1,ft2,ft3,ft4}. For other modified
theories which belong to that class we referee the reader to
\cite{ff1,ff2,ff4,ff5,ff6,ff7,ff8,ff9,ff10,ff11,ff12,ff13,ff14,ff15,ff16} and
references therein.

In this work we are interested in $f\left(  R\right)  $-gravity in the metric
formalism \cite{Buda}, where Action Integral in a four-dimensional manifold is
given by the expression $S=\int dx^{4}\sqrt{-g}f\left(  R\right)  $. In this
theory, variable $R$ corresponds to the Ricci scalar of the underlying
geometry with line element $g_{\mu\nu}$; consequently, General Relativity with
or without the cosmological constant is fully recovered when $f\left(
R\right)  $ is a linear function. Various specific functional forms of
$f\left(  R\right)  -$theory have been proposed in the literature in order to
describe the various phases of the universe. The quadratic model $f\left(
R\right)  =R+\alpha R^{2}$ can describe well the inflationary era of our
universe \cite{sta1,bot}. The natural extension of the latter inflationary
model is the $f\left(  R\right)  =R+\alpha R^{n}$ model \cite{v11} which
provides power-law attractors. For other $f\left(  R\right)  $-models\ with
applications in the late acceleration phase of the universe see
\cite{rev1,rev2,rev3,rev4} and references therein.

$f\left(  R\right)  -$theory is a fourth-order theory and it is dynamical
equivalent to the Brans-Dicke theory with zero value for the Brans-Dicke
parameter. The scalar field attributes the extra degree of freedom such that
the theory is written as a second-order theory but with extra dependent
variables so that the total degrees of freedom are the same. The theory is
nonlinear and there are few exact solutions, either for spacetimes with one
free function such as the Friedmann--Lema\^{\i}tre--Robertson--Walker metric
(FLRW) which is usually applied in modern cosmology. Indeed, in the case of a
spatially flat FLRW spacetime the de Sitter solution, $R=R_{0}$, is recovered
when there exists a solution to the algebraic equation $R_{0}f^{\prime}\left(
R_{0}\right)  =2f\left(  R_{0}\right)  ~$\cite{bot}. In addition, power law
solutions, which describe an ideal gas with constant equation parameter, are
recovered when $f\left(  R\right)  =f_{0}R^{n}$,~$n\neq0,1,2$. However, the
latter exact solutions do not describe the generic analytic solution for the
corresponding field equations because they are valid only for specific initial
conditions. Some analytic solutions have been found by searching for
conservation laws for the field equations and making a conclusion about the
integrability of the\ gravitational model by writing the analytic solution
with the use of closed-form functions or making use of theorems from the
theory of Analytic Mechanics, for instance see \cite{ns1,ns2,ns4,ns5}.

We focus on the determination of exact solutions of the Wheeler-DeWitt (WdW)
equation \cite{wd1} in $f\left(  R\right)  $-cosmology. The WdW equation is
mainly applied in quantum cosmology. \ Recall that in modern cosmology we
assume that the spacetime is described by the FLRW metric with zero spatial
curvature. WdW is an equation of Klein-Gordon type, where the dependent
variable is denoted to describe the wavefunction of the universe and the
independent variables are the dynamical variables of the classical system.
There are various issues such that there is not a unique way for one to define
probability \cite{wdw1,wdw2}. Also there is the so-called problem of time,
because time is involved in the wavefunction through the dynamical variables
\cite{wdw4,wdw5,wdw6,wdw7}.

A previous analysis of the exact solutions of WdW in $f\left(  R\right)
$-cosmology was published in \cite{vak1,ss1,ss2}. Specifically in \cite{vak1}
there was found that for the special power-law theory $f\left(  R\right)
=R^{\frac{3}{2}}\,,$ the classical solution can be recovered from the solution
of the WdW equation. The case $f\left(  R\right)  =R^{\frac{3}{2}}$ describes
an integrable cosmological model which admits a conservation law linear in the
momentum. That approach has been extended and applied in other gravitational
models, such as anisotropic universes \cite{wdw2}, static spherical symmetric
spacetimes \cite{bv01,tch1,tch4}, inhomogeneous spacetimes \cite{tch2} and
electromagnetic three-dimensional pp-wave spacetimes \cite{tch3}. \ 

In our consideration we determine a family of Lie-B\"{a}cklund transformations
for the WdW equation for some specific models of $f\left(  R\right)
$-cosmology. The models of $f\left(  R\right)  $-cosmology that we study form
integrable dynamical systems where the conservation laws which ensure the
integrability are constructed by point transformations which leave the
variational integral invariant. The plan of the paper is as follows.

In Section \ref{sec2} we present the basic equations of $f\left(  R\right)
$-cosmology. The main mathematical materials necessary for the analysis of the
present work are given in Section \ref{sec3}. Specifically, we show how
Lie-B\"{a}cklund transformations can be constructed for the conformally
invariant Klein-Gordon equation by using the point symmetries of the classical
Hamiltonian system. In addition we show how the Lie-B\"{a}cklund operators are
applied in order to determine similarity solutions for the WdW equation. The
context of the one-dimensional optimal system is discussed. Section \ref{sec4}
includes the main material of our analysis. For four integrable classical
models of $f\left(  R\right)  $-cosmology we write the WdW equation and we
determine the infinitesimal generators of the point transformations where the
WdW equation is invariant. From the infinitesimal generators we construct
the~Lie-B\"{a}cklund operators and we find the similarity solutions. In order
our results to be completed the one-dimensional optimal system is determined
for each model. For the models of our study we observe that the classical
limit is always recovered. Finally in Section \ref{sec5}, we discuss our
results and we draw our conclusions.

\section{$f\left(  R\right)  $-Cosmology}

\label{sec2}

For a spatially flat FLRW background space with line element%
\begin{equation}
ds^{2}=-dt^{2}+a^{2}\left(  t\right)  \left(  dx^{2}+dy^{2}+dz^{2}\right)  ,
\end{equation}
and Ricciscalar%
\begin{equation}
R=6\left(  \dot{H}+2H\right)  ~, \label{fr.05}%
\end{equation}
the gravitational field equations of $f\left(  R\right)  $-gravity are
calculated to be \cite{fr1}
\begin{equation}
3f^{\prime}H^{2}=\frac{f^{\prime}R-f}{2}-3Hf^{\prime\prime}\dot{R},
\label{fr.06}%
\end{equation}%
\begin{equation}
2f^{\prime}\dot{H}+3f^{\prime}H^{2}=-2Hf^{\prime\prime}\dot{R}-\left(
f^{\prime\prime\prime}\dot{R}^{2}+f^{\prime\prime}\ddot{R}\right)
-\frac{f-Rf^{\prime}}{2}, \label{fr.07}%
\end{equation}
where $H\left(  t\right)  $ is the Hubble function, $H\left(  t\right)
=\frac{\dot{a}\left(  t\right)  }{a\left(  t\right)  },$ dot indicates
derivative with respect to the independent variable \textquotedblleft%
$t$\textquotedblright; and a prime denotes derivative with respect to the
Ricciscalar, that is, $f^{\prime}\left(  R\right)  =\frac{df\left(  R\right)
}{dR}$.

The latter field equations can be written in an equivalently form as follows
\cite{fr1}%
\begin{equation}
G_{~\nu}^{\mu}=k_{eff}T_{f~~\nu}^{\mu} \label{fr.07a}%
\end{equation}
where\ now $G_{\nu}^{\mu}$, is the Einstein tensor, $k_{eff}$ is a varying
"Einstein-constant" defined as $k_{eff}=\frac{1}{f^{\prime}\left(  R\right)
}$, and~$T_{f~~\nu}^{\mu}$ is the effective energy momentum tensor which
attributes the geometrodynamical degrees of freedom of the higher-order of
gravity. Indeed, the energy-momentum tensor$~T_{f~~\nu}^{\mu}$ $\ $is defined
as%
\[
T_{\mu\nu}=\left(  \rho_{f}+p_{f}\right)  u_{\mu\nu}+p_{f}g_{\mu\nu},
\]
where the energy density $\rho_{f}$ and pressure term $p_{f}$ are defined as
\cite{fr1}
\begin{equation}
\rho_{f}=\frac{f^{\prime}R-f}{2}-3Hf^{\prime\prime}\dot{R}, \label{fr.11}%
\end{equation}%
\begin{equation}
p_{f}=2Hf^{\prime\prime}\dot{R}+\left(  f^{\prime\prime\prime}\dot{R}%
^{2}+f^{\prime\prime}\ddot{R}\right)  +\frac{f-Rf^{\prime}}{2}. \label{fr.12}%
\end{equation}

Hence, the field equations are%
\begin{equation}
3H^{2}=k_{eff}\rho_{f}~,~2\dot{H}+3H^{2}=-k_{eff}p_{f}, \label{fr.09}%
\end{equation}
while the equation of state parameter for the effective fluid%
\begin{equation}
w_{f}=\frac{p_{f}}{\rho_{f}}=-\frac{\left(  f-Rf^{\prime}\right)
+4Hf^{\prime\prime}\dot{R}+2\left(  f^{\prime\prime\prime}\dot{R}%
^{2}+f^{\prime\prime}\ddot{R}\right)  }{\left(  f-Rf^{\prime}\right)
+6Hf^{\prime\prime}\dot{R}}, \label{fr.12a}%
\end{equation}
Note that the latter expression for $f\left(  R\right)  =R-2\Lambda$ gives
$w_{f}=-1$ which means that the theory of General Relativity with the
cosmological constant is recovered.

\subsection{Minisuperspace approach}

The gravitational field equations (\ref{fr.05}), (\ref{fr.06}) and
(\ref{fr.07}) can be derived by a variation principle of the Action integral%
\begin{equation}
A=\int L\left(  N,a,\dot{a},R,\dot{R}\right)  dadRdN \label{fr.12b}%
\end{equation}
where $L\left(  N,a,\dot{a},R,\dot{R}\right)  $ is defined as \cite{ns2}%

\begin{equation}
L\left(  N,a,\dot{a},R,\dot{R}\right)  =\frac{1}{N}\left(  6af^{\prime}\dot
{a}^{2}+6a^{2}f^{\prime\prime}\dot{a}\dot{R}\right)  +Na^{3}\left(  f^{\prime
}R-f\right)  \label{fr.13}%
\end{equation}
where $N\left(  t\right)  $ is a generic lapse function for the\ FLRW metric,
such that the Hubble function is defined $H\left(  t\right)  =\frac{1}{N}%
\frac{\dot{a}}{a}$. We note that Lagrangian (\ref{fr.13}) is a singular
Lagrangian since $\frac{\partial L}{\partial\dot{N}}=0$. Lagrangian
(\ref{fr.13}) defines a constraint system, with constraint equation
$\frac{\partial L}{\partial N}=0$. The two-second order order differential
equations follow by the variation with respect to the variables $\left\{
a,R\right\}  $, that is, $\frac{d}{dt}\frac{\partial L}{\partial\dot{a}}%
-\frac{\partial L}{\partial a}=0$ and $\frac{d}{dt}\frac{\partial L}%
{\partial\dot{R}}-\frac{\partial L}{\partial R}=0$.

Lagrangian (\ref{fr.13}) is of the form%
\begin{equation}
L\left(  N,a,\dot{a},R,\dot{R}\right)  =\frac{1}{2N}\mathcal{G}_{AB}%
\frac{dq^{A}}{dt}\frac{dq^{B}}{dt}-N\mathcal{U}({q}^{C}\mathbf{)}%
\end{equation}
where $\ q^{A}=\left(  a,R\right)  $,~$U\left(  q^{C}\right)  =-a^{3}\left(
f^{\prime}R-f\right)  $ and~$\mathcal{G}_{AB}$ is the minisuperspace defined
as
\begin{equation}
\mathcal{G}_{AB}=%
\begin{pmatrix}
12af^{\prime} & 6a^{2}f^{\prime\prime}\\
6a^{2}f^{\prime\prime} & 0
\end{pmatrix}
.\label{s1}%
\end{equation}
The second-rank tensor $\mathcal{G}_{AB}$ defines the space where the
dynamical variables $\left\{  a,R\right\}  $ evolve.

We can define a canonical momenta for the variable $\left\{  a,R\right\}  $,
and write the point-like Lagrangian (\ref{fr.13}) as a Hamiltonian system. It
follows that the two momentum $p_{a}=\frac{\partial L}{\partial a}$, and
$p_{R}=\frac{\partial L}{\partial\dot{R}}$ are
\begin{equation}
Np_{a}=12af^{\prime}\dot{a}+6a^{2}f^{\prime\prime}\dot{R}~,~Np_{R}%
=6a^{2}f^{\prime\prime}\dot{a}%
\end{equation}
so the Hamiltonian function is
\begin{equation}
\mathcal{H}=N\left[  \frac{p_{a}p_{R}}{6a^{2}}-\frac{f^{\prime}p_{R}^{2}%
}{6a^{3}}-a^{3}\left(  f^{\prime}R-f\right)  \right]  , \label{ac.17}%
\end{equation}
or equivalently%
\begin{equation}
\mathcal{H}=N\left(  \frac{1}{2}\mathcal{G}^{AB}P_{A}P_{B}+\mathcal{U}({q}%
^{C}\mathbf{)}\right)  .
\end{equation}
where $P_{A}=\left(  p_{a},p_{R}\right)  $ is the canonical momentum.

Hence, the constraint equation provides
\begin{equation}
\mathcal{H}\left(  a,R,p_{a},p_{R}\right)  \equiv0, \label{ac.18}%
\end{equation}
while the rest of the field equations are given by the Hamilton equations%
\begin{equation}
\dot{a}=\frac{\partial\mathcal{H}}{\partial p_{a}}~,~\dot{R}=\frac
{\partial\mathcal{H}}{\partial p_{R}}, \label{ac.19}%
\end{equation}%
\begin{equation}
\dot{p}_{a}=-\frac{\partial\mathcal{H}}{\partial a}~,~\dot{p}_{R}%
=-\frac{\partial\mathcal{H}}{\partial R}, \label{ac.20}%
\end{equation}
that is%
\begin{equation}
\frac{1}{N}\dot{a}=\frac{p_{R}}{6a^{2}}~\ ,~\frac{1}{N}\dot{R}=\frac{p_{a}%
}{6a^{2}}-\frac{f^{\prime}p_{R}}{3a^{3}} \label{ac.21}%
\end{equation}%
\begin{equation}
\frac{1}{N}\dot{p}_{a}=-\frac{p_{a}p_{R}}{3a^{3}}+\frac{f^{\prime}p_{R}^{2}%
}{2a^{4}}+3a^{2}\left(  f^{\prime}R-f\right)  , \label{ac.22}%
\end{equation}
and\qquad%
\begin{equation}
\frac{1}{N}\dot{p}_{R}=\frac{f^{\prime\prime}p_{R}^{2}}{6a^{3}}+a^{3}%
f^{\prime\prime}R. \label{ac.23}%
\end{equation}

\subsection{Wheeler-DeWitt equation}

Constraint equation (\ref{ac.18}) yields the WdW equation $\hat{H}%
\Psi(\mathbf{q})=0$, where $\hat{H}$ is the Hamiltonian operator under
canonical quantization, $P_{A}=\frac{1}{\sqrt{G}}\frac{\partial}{\partial
q^{A}}$.

The operator $\hat{H}$ is defined as \cite{Wil}
\begin{equation}
\hat{H}\Psi(\mathbf{q})=\left(  \frac{1}{2}\Delta_{L}+\mathcal{U}%
(\mathbf{q)}\right)  \Psi(\mathbf{q})\equiv0, \label{ac.24}%
\end{equation}
in which $\Delta_{L}$ is the conformal Laplace operator defined as
\begin{equation}
\Delta_{L}=\Delta+\frac{n-2}{4(n-1)}\mathcal{R},
\end{equation}
where $\mathcal{R}$ is the Ricciscalar of the minisuperspace $\mathcal{G}%
_{AB}$ and $n=\dim\mathcal{G}_{AB}$ and $\Delta$ is the Laplace operator, that
is,%
\begin{equation}
\Delta=\frac{1}{\sqrt{-\mathcal{G}}}\partial_{A}\left(  \sqrt{-\mathcal{G}%
}\mathcal{G}^{AB}\partial_{B}\right)  .
\end{equation}

For the second-rank tensor (\ref{s1}) we calculate $n=2$, which means that
$\Delta_{L}=\Delta$. The conformal Laplace operator $\Delta_{L}$ has the
property that it is invariant under conformal transformations, such a
requirement it is necessary in the case of quantum cosmology since the theory
should be conformal invariant because of the arbitance of the lapse function
$N\left(  t\right)  $. While in the case where $n=2$ operator $\Delta_{L}$
follows from the canonical quantization $P_{A}\simeq\frac{1}{\sqrt{G}}%
\frac{\partial}{\partial q^{A}}$, for higher-dimensional spaces, $n\geq3$, the
conformal Laplace operator $\Delta_{L}$ follows from the canonical
quantization only for conformally flat spaces, and in general the term
$\frac{n-2}{4(n-1)}\mathcal{R}$ should be added by hand. However, which
quantization process which provides the operator $\Delta_{L}$ for \ $n\geq3$
from a point-Lie Hamiltonian function is still an open problem.

In general and in terms of the $1+3$ decomposition notation of GR the WdW
equation it follows from the Hamiltonian constraint%

\begin{equation}
\mathcal{H}\Psi=\left[  -4\kappa^{2}\mathcal{G}_{ijkl}\frac{\delta^{2}}{\delta
h_{ij}\delta h_{kl}}+\frac{\sqrt{h}}{4\kappa^{2}}\left(  -\mathcal{R}%
+2\Lambda+4\kappa^{2}T^{00}\right)  \right]  \Psi=0, \label{WDW1}%
\end{equation}
where~$\mathcal{G}_{ijkl}$ is defined as
\begin{equation}
\mathcal{G}_{ijkl}=\frac{1}{2\sqrt{h}}\left(  h_{ik}h_{jl}+h_{il}h_{jk}%
-h_{ij}h_{kl}\right)  , \label{WDW2}%
\end{equation}
is the the metric of superspace, the space of all 3-geometries with metric
$h_{ij}$ and Ricci scalar $\mathcal{R}$, and the matter configuration.

We note that in general the WdW equation (\ref{WDW1}) is a hyperbolic
functional differential equation on superspace, where in the case of the
minisuperspace approximation it is reduced to a single equation for all the
points of the superspace.

As far as our model of $f\left(  R\right)  $-cosmology is concerned, with
constraint equation (\ref{ac.17}), the WdW equation in the minisuperspace
approach is found to be%
\begin{equation}
\frac{1}{a^{2}f^{\prime\prime}}\Psi_{,aR}-\frac{f^{\prime}}{a^{3}\left(
f^{\prime\prime}\right)  ^{2}}\Psi_{,RR}+\left(  \frac{f^{\prime}%
f^{\prime\prime\prime}}{a^{3}\left(  f^{\prime\prime}\right)  ^{3}}-\frac
{1}{a^{3}f^{\prime\prime}}\right)  \Psi_{,R}-6a^{3}\left(  f^{\prime
}R-f\right)  \Psi=0. \label{wdw.01}%
\end{equation}

For the latter linear second-order partial differential equation we shall
determine exact solutions for specific forms of $f\left(  R\right)  $
function. In the following section we present the main mathematical tools
which will be applied in order to determine solutions for equation
(\ref{wdw.01}).

\section{Constructing similarity solutions}

\label{sec3}

Consider the partial differential equation $H\left(  q^{A},\Psi,\Psi_{,A}%
,\Psi_{,AB},...\right)  \equiv0$ where $q^{A}=\left(  q^{1},q^{2}%
,...,q^{n}\right)  $ denotes the $n-$independent variables and $\Psi
=\Psi\left(  q^{A}\right)  $ is the dependent variable. Let the differential
equation $H\left(  q^{A},\Psi,\Psi_{,A},\Psi_{,AB},...\right)  $ be invariant
under the infinitesimal one-parameter point transformation $q^{n}\rightarrow
q^{n}+\varepsilon$, \ then the differential equation can be rewritten as
$\bar{H}\left(  q^{\alpha},\Psi,\Psi_{,\alpha},\Psi_{,\alpha\beta},...\right)
$, where $\Psi=\Psi\left(  q^{\alpha}\right)  $ and $q^{a}=\left(  q^{1}%
,q^{2},...,q^{n-1}\right)  $. This process is called similarity transformation
or similarity reduction, while the solutions which follow by that kind of
transformations are called similarity solutions.

When the differential equation $H\left(  q^{A},\Psi,\Psi_{,A},\Psi
_{,AB},...\right)  $ is invariant under the infinitesimal one-parameter point
transformation $q^{n}\rightarrow q^{n}+\varepsilon$, then we shall say that
the differential equation admits the Lie point symmetry $X=\partial_{q^{n}}$
and vice versa.

In general, the differential equation $H\left(  q^{A},\Psi,\Psi_{,A}%
,\Psi_{,AB},...\right)  $ is invariant under the infinitesimal one-parameter
point transformation
\begin{equation}
q^{A}\rightarrow q^{A}+\varepsilon\xi^{A}\left(  q^{B},\Psi\right)
~,~\Psi\rightarrow\Psi+\varepsilon\eta\left(  q^{B},\Psi\right)  \label{ls.01}%
\end{equation}
if and only if there exists a function $\lambda\left(  q^{B},\Psi\right)  $
such that$\xi^{{}}$%
\begin{equation}
\left[  X^{\left[  k\right]  },H\right]  =\lambda H, \label{ls.02}%
\end{equation}
where $X^{\left[  k\right]  }$ is the $k$th extension of the vector field
$X=\xi^{A}\partial_{A}+\eta\partial_{\Psi}$ in the jet-space $\left\{
q^{A},\Psi,\Psi_{,A},\Psi_{,AB},...\right\}  $. The transformation
$q^{A}\rightarrow\bar{q}^{A}\left(  q^{B}\right)  $ which transforms the
generic field $X=\xi^{A}\left(  q^{B},\Psi\right)  \partial_{A}+\eta\left(
q^{B},\Psi\right)  \partial_{\Psi}$ in the form $X=\partial_{q^{n}}$ is called
canonical transformation.

The Lie symmetries for the conformal invariant Klein-Gordon equation
(\ref{ac.24}) have been studied before in \cite{schan}. Specifically it has
been found that the generic Lie symmetry has the form%
\begin{equation}
X=\xi^{A}\left(  q^{B}\right)  \partial_{A}+\left(  \frac{2-n}{2}\psi\left(
q^{A}\right)  \Psi+a_{0}\Psi+b\left(  q^{A}\right)  \right)  \partial_{\Psi},
\label{ls.03}%
\end{equation}
in which $a_{0}$ is a constant, $b\left(  q^{A}\right)  $ is a solution of the
original equation (\ref{ac.24}) and represents the infinity number of
solutions, since the equation is linear, and $\xi^{A}\left(  q^{A}\right)  $
is a conformal vector field for the minisuperspace $G_{AB}\left(
q^{C}\right)  $, with conformal factor $\psi\left(  q^{B}\right)  $, that is,
\[
\mathcal{L}_{\xi}G_{AB}\left(  q^{C}\right)  =2\psi\left(  q^{C}\right)
G_{AB}\left(  q^{C}\right)  ,
\]
$\mathcal{L}_{\xi}$ denotes the Lie derivative with respect to the vector
field $\xi$.

In addition, the conformal vector field and the potential $\mathcal{U}%
(q^{C}\mathbf{)}$ satisfy the constraint condition%
\begin{equation}
\mathcal{L}_{\xi}\mathcal{U}(q^{C}\mathbf{)}+2\psi\left(  q^{C}\right)
\mathcal{U}(q^{C}\mathbf{)=}0. \label{ls.04}%
\end{equation}
\bigskip

By definition, if $X=\xi^{A}\left(  q^{B},\Psi\right)  \partial_{A}%
+\eta\left(  q^{B},\Psi\right)  \partial_{\Psi}$ is a Lie point symmetry for
the differential equation $H\left(  q^{A},\Psi,\Psi_{,A},\Psi_{,AB}%
,...\right)  $, the symmetry vector $\hat{X}=\left(  \eta\left(  q^{B}%
,\Psi\right)  -\xi^{A}\left(  q^{B},\Psi\right)  \Psi_{,A}\right)
\partial_{\Psi}$ is a Lie-B\"{a}cklund symmetry. Vector field $\hat{X}$ is the
canonical form of the vector field $X$.

A Lie-B\"{a}cklund symmetry preserves the set of solutions for the
differential equation, that is,%
\begin{equation}
\hat{X}\left(  \Psi\right)  =\lambda_{0}\Psi~,~\lambda_{0}=cons\not t  .
\label{ls.05}%
\end{equation}
Condition (\ref{ls.05}) provides a constraint equation which will be used in
the following to solve the WdW equation (\ref{wdw.01}).

The symmetry condition (\ref{ls.04}) where $\xi^{A}\left(  q^{B}\right)  $ is
a conformal vector field of the minisuperspace has been found before in
\cite{ns1} for the variational symmetries for singular Lagrangians of the form
of (\ref{fr.13}). Indeed, for every variational symmetry of (\ref{fr.13}) a
Lie point symmetry and consequently a Lie-B\"{a}cklund can be constructed for
the WdW equation (\ref{wdw.01}). However, that it is not the only relation
between variational symmetries of classical Lagrangians and Lie symmetries of
the WdW equation.

If we consider that the lapse-function $N\left(  t\right)  $ in the Lagrangian
(\ref{fr.13}) is fixed, then we can apply the results for the variational
symmetries of regular Lagrangians \cite{ns2}. As we shall see for the
$f\left(  R\right)  $-theory we recover the results of \cite{ns111} while also
we found new Lie-B\"{a}cklund operators which will be used to determine new
similarity solutions for equation (\ref{wdw.01}).

However, the Lie point symmetries of regular systems can be time-dependent,
something which is not true for the WdW equation. Below we show two cases of
important interest where we show how Lie-B\"{a}cklund operators are
constructed by using the time-dependent symmetries of the regular Lagrangian.

\subsection{Higher-order Lie-B\"{a}cklund operators}

We show, for two models of special interest, how to construct Lie-B\"{a}cklund
operators for the conformal Laplace equation (\ref{ac.24}) by using
time-dependent point symmetries of regular Lagrangians.

\subsubsection{Oscillator}

Consider the point-like regular Lagrangian%

\begin{equation}
L=\frac{1}{2}\left(  \dot{x}^{2}+h_{AB}\left(  y^{C}\right)  \dot{y}^{A}%
\dot{y}^{B}\right)  +\frac{1}{2}\mu^{2}x^{2}+F\left(  y^{C}\right)  .
\label{CR.02}%
\end{equation}
where $h_{AB}\left(  y^{C}\right)  $ and $F\left(  y^{C}\right)  $ are
arbitrary functions.

Lagrangian (\ref{CR.02}) admits the variational symmetries, $\partial_{t}~$and
$e^{\pm\mu t}\partial_{x}$, with respective gauge function $f\left(
t,x,y^{A}\right)  =\mu e^{\pm}x$. Consequently, from the two-latter
variational symmetries for the dynamical system with Lagrangian (\ref{CR.02})
we can construct the time-dependent conservation laws
\begin{equation}
I_{\pm}=e^{\pm\mu t}\dot{x}\mp\mu e^{\pm\mu t}x.
\end{equation}
It is easy to show that the combined integral $I_{0}=I_{+}I_{-}$ is time
independent and equals%
\begin{equation}
I_{0}=\dot{x}^{2}-\mu^{2}x^{2}. \label{CR.04}%
\end{equation}

The corresponding conformal invariant Klein-Gordon equation is%
\begin{equation}
\Psi_{xx}+h^{AB}\left(  y^{C}\right)  \Psi_{A}\Psi_{B}-\Gamma^{A}\left(
y^{C}\right)  \Psi_{A}+\frac{n-2}{4\left(  n-1\right)  }R\left(  y^{C}\right)
\Psi-\mu^{2}x^{2}\Psi-F\left(  y^{C}\right)  \Psi=0. \label{CR.06}%
\end{equation}
Equation (\ref{CR.06}) does not admit any Lie point symmetry for general
$h_{AB},$ $F\left(  y^{C}\right)  ~$while $R\left(  y^{C}\right)  $ is the
Ricciscalar for the metric $h_{AB}$.~

We observe that equation (\ref{CR.06}) is separable with respect to $x$.
Indeed the solution can be written in the form $\Psi\left(  x,y^{A}\right)
=w\left(  x\right)  S\left(  y^{A}\right)  .$ This implies that the operator
\begin{equation}
\hat{I}=D_{x}D_{x}-\mu^{2}x^{2}-I_{0}%
\end{equation}
satisfies $\hat{I}\Psi=\bar{I}_{0}\Psi,$ where $D_{i}$ is the operator
$D_{i}=\partial_{A}+\Psi_{A}\partial_{\Psi}+\Psi_{AB}\partial_{\Psi_{A}}+....$

From the latter it follows that the Klein Gordon equation (\ref{CR.06})
possesses a Lie-B\"{a}cklund symmetry with generating vector
\begin{equation}
\hat{X}=\left(  \Psi_{xx}-\mu^{2}x^{2}\Psi\right)  \partial_{\Psi}.
\end{equation}

\subsubsection{Ermakov-Pinney$~$system}

The second case we consider is that of the Ermakov-Pinney system. Let us
assume the generic regular Lagrangian function%
\begin{equation}
L=\frac{1}{2}\left(  \dot{r}^{2}+r^{2}h_{AB}\left(  y^{C}\right)  \dot{y}%
^{A}\dot{y}^{B}\right)  +\frac{1}{2}\mu^{2}r^{2}-\frac{F\left(  y^{C}\right)
}{r^{2}} \label{CR.07}%
\end{equation}
where $h_{AB}\left(  y^{C}\right)  $ and $F\left(  y^{C}\right)  $ are
arbitrary functions.

The dynamical system described by the Lagrangian (\ref{CR.07}) admits the
time-dependent conservation laws%
\begin{align}
I_{+}  &  =\frac{h}{\mu}e^{2\mu t}-e^{2\mu s}r\dot{r}+\mu e^{2\mu t}%
r^{2}\label{GERSN.5}\\
I_{-}  &  =\frac{h}{\mu}e^{-2\mu t}+e^{-2\mu t}r\dot{r}+\mu e^{-2\mu t}r^{2}.
\label{GERSN.6}%
\end{align}
where $h$ is the value for the integral of motion described by the Hamiltonian
for Lagrangian (\ref{CR.07}).

In a similar way as before we construct the autonomous first integral
\cite{eer1}
\begin{equation}
\Phi_{0}=h^{2}-I_{+}I_{-},
\end{equation}
which equals%
\begin{equation}
\Phi_{0}=r^{4}h_{DB}\dot{y}^{A}\dot{y}^{B}+2F\left(  y^{C}\right)  .
\label{Er22}%
\end{equation}
This is the well known Ermakov invariant, also known as Lewis invariant.

Consider now the conformal invariant Klein-Gordon equation%
\begin{equation}
\Psi_{rr}+\frac{1}{r^{2}}h^{AB}\Psi_{AB}+\frac{n-1}{r}\Psi_{r}-\frac{1}{r^{2}%
}\Gamma^{A}\Psi_{A}+\frac{n-2}{4\left(  n-1\right)  }\frac{1}{r^{2}}R\left(
y^{C}\right)  \Psi+\mu^{2}r^{2}\Psi+\frac{1}{r^{2}}F\left(  y^{C}\right)
\Psi=0, \label{CR.11}%
\end{equation}
where $R\left(  y^{C}\right)  $ is the Ricciscalar of the metric
$h_{AB}\left(  y^{C}\right)  .$ The latter equation does not have any Lie
point symmetries.

However, the latter Klein-Gordon equation is separable, in the sense that
$\Psi\left(  r,y^{C}\right)  =w\left(  r\right)  S\left(  y^{C}\right)  .$
Then we shall say that the operator%
\begin{equation}
\hat{\Phi}=h^{AB}D_{A}D_{B}-\Gamma^{A}D_{A}+F\left(  y^{C}\right)  +\frac
{n-2}{4\left(  n-1\right)  }R\left(  y^{C}\right)  -\Phi_{0},
\end{equation}
satisfies the equation $\hat{\Phi}\Psi=0$ which means that the Klein Gordon
equation (\ref{CR.11}) admits the Lie-B\"{a}cklund symmetry with generator
\begin{equation}
\bar{X}=\left(  h^{AB}D_{A}D_{B}\Psi-\Gamma^{A}D_{A}\Psi+F\left(
y^{C}\right)  \Psi+\frac{n-2}{4\left(  n-1\right)  }R\left(  y^{C}\right)
\Psi\right)  \partial_{\Psi}.
\end{equation}

\subsection{One-dimensional optimal system}

However, Lie symmetries are used to find new similarity solutions for other
similarity solutions by applying the adjoint representation of the admitted
Lie group for the given differential equations. Hence, it is important to
determine the one-dimensional optimal system for the admitted Lie algebra for
the equation of our study. In that case we will determine all the unique
similarity solutions which can not derived by adjoint transformation. In the
following we give the definition of the adjoint operator as also when two Lie
point symmetries are connected through the adjoint representation.

Let a given differential equation $H\left(  q^{A},\Psi,\Psi_{,A},\Psi
_{,AB},...\right)  $ to admit a $n$-dimensional Lie algebra $G_{n}$ with
elements $X_{1},~X_{2},~...~X_{n}$. Then we shall say that the two vector
fields \cite{olver,kumei}
\begin{equation}
Z=\sum\limits_{i=1}^{n}a_{i}X_{i}~,~W=\sum\limits_{i=1}^{n}b_{i}%
X_{i}~,~\text{\ }a_{i},~b_{i}\text{ are constants,} \label{sw.04}%
\end{equation}
are equivalent if and only if $W=lim_{j=i}^{n}Ad\left(  \exp\left(
\varepsilon_{i}X_{i}\right)  \right)  Z\mathbf{~}$or $~\frac{b_{i}}{a_{i}%
}=c~,~c=const,~$\ in which$~Ad\left(  \exp\left(  \varepsilon_{i}X_{i}\right)
\right)  $ is the adjoint operator defined as%
\begin{equation}
Ad\left(  \exp\left(  \varepsilon X_{i}\right)  \right)  X_{j}=X_{j}%
-\varepsilon\left[  X_{i},X_{j}\right]  +\frac{1}{2}\varepsilon^{2}\left[
X_{i},\left[  X_{i},X_{j}\right]  \right]  +...~. \label{sw.07}%
\end{equation}

\section{Similarity solutions of the Wheeler-DeWitt equation}

\label{sec4}

As we discussed before, in order to solve the WdW equation (\ref{wdw.01}) we
will construct differential operators by using the variational symmetries for
the classical system. Such an analysis was performed before in \cite{ns2}
where the unknown function $f\left(  R\right)  $ which defines the theory is
constrained by the requirement the field equations in $f\left(  R\right)
$-cosmology to admit conservation laws generated by point symmetries.

In Lagrangian (\ref{fr.13}) we assume that $N\left(  t\right)  =1$. Therefore,
for arbitrary function $f\left(  R\right)  $ the dynamical system is
autonomous and admits the point symmetry $\partial_{t}$. However ,the latter
symmetry does not provide any differential operator for the WdW equation
(\ref{wdw.01}).

In addition, there are four specific functions of $f\left(  R\right)
$-function where Lagrangian (\ref{fr.13}) is transformed such that the
variation of the Action Integral (\ref{fr.12b}) to be invariant. Specifically,
the cases we shall study are (A) $f\left(  R\right)  =R^{\frac{3}{2}};~$(B)
$f\left(  R\right)  =R^{\frac{7}{8}};~$(C) $f\left(  R\right)  =\left(
R-2\Lambda\right)  ^{\frac{3}{2}}$ and (D) $f\left(  R\right)  =\left(
R-2\Lambda\right)  ^{\frac{7}{8}}$. \ The first two models are power-law
models; however models D and E, belong to a family of models which are called
$\Lambda_{bc}$CDM with general form$~f\left(  R\right)  =\left(
R^{b}-2\Lambda\right)  ^{c}$ \cite{ff1}. Indeed, model D$~$is the
$\Lambda_{1\frac{3}{2}}$CDM while model E corresponds to the $\Lambda
_{1\frac{7}{8}}$CDM.

At this point it is important to mention that because the WdW equation
(\ref{wdw.01}) is a linear second-order partial differential equation, it
admits for arbitrary function $f\left(  R\right)  $ the two symmetry vectors
$X_{\Psi}=\Psi\partial_{\Psi}$ and $X_{b}=b\left(  a,R\right)  \partial_{\Psi
}$, in which $b\left(  a,R\right)  $ is a solution of (\ref{wdw.01}). Symmetry
$X_{b}$ denotes the infinity number of solutions for the partial differential
equations. However, $X_{b}$ plays no role in the derivation of similarity
solutions and for that reason we will omit it.

\subsection{Case A: Power law model $R^{\frac{3}{2}}$}

For the first model of our consideration, with $f\left(  R\right)
=R^{\frac{3}{2}}\,$, the point-like Lagrangian of the classical field
equations becomes%
\begin{equation}
L\left(  a,\dot{a},R,\dot{R}\right)  =9a\sqrt{R}\dot{a}^{2}+\frac{9a^{2}%
}{2\sqrt{R}}\dot{a}\dot{R}+\frac{a^{3}}{2}R^{\frac{3}{2}}. \label{FR.32}%
\end{equation}
However, under the change of coordinates $\left\{  a,R\right\}  \rightarrow
\left\{  z,w\right\}  $ with the relation $a=\left(  \frac{9}{2}\right)
^{-\frac{1}{3}}\sqrt{z}\;,\;R=\frac{w^{2}}{z}~$the point-like Lagrangian
(\ref{FR.32}) is simplified as follows,
\begin{equation}
L\left(  z,w,\dot{z},\dot{w}\right)  =\dot{z}\dot{w}+\frac{1}{9}w^{3}
\label{WD.00}%
\end{equation}

Consequently, the field equations in the Hamiltonian formalism become
\begin{equation}
H=p_{z}p_{w}-\frac{1}{9}w^{3}\equiv0 \label{WD.01}%
\end{equation}%
\begin{equation}
\dot{z}=p_{w}~~~\dot{w}=p_{z}~,~\dot{p}_{z}=0~~~~\dot{p}_{w}=\frac{1}{3}w^{2}.
\end{equation}
The latter system can be easily integrated and the exact solution is presented
in \cite{ns2}.

From the Hamiltonian (\ref{WD.01}) results the WdW equation
\begin{equation}
\Psi_{zw}-\frac{1}{9}w^{3}\Psi=0. \label{WD.02}%
\end{equation}
which admit the Lie point symmetries
\begin{equation}
X_{1}=\partial_{z}~,~X_{2}=\frac{1}{w^{3}}\partial_{w}~,~X_{3}=z\partial
_{z}-\frac{w}{4}\partial_{w}~,~X_{\Psi}=\Psi\partial_{\Psi},
\end{equation}
or in canonical form the Lie-B\"{a}cklund operators%
\begin{equation}
\hat{X}_{1}=\Psi_{z}\partial_{\Psi},~\hat{X}_{2}=\frac{1}{w^{3}}\Psi
_{w}\partial_{\Psi},~\hat{X}_{3}=\left(  z\Psi_{z}-\frac{w}{4}\Psi_{w}\right)
\partial_{\Psi}~,~\hat{X}_{\Psi}=\Psi\partial_{\Psi}%
\end{equation}
The commutators and the Adjoint representation of the admitted Lie algebra are
presented in tables \ref{tabl1} and \ref{tabl2}.

Therefore, from the Adjoint representation we determine the one-dimensional
optimal system%
\begin{align*}
&  \left\{  X_{1}\right\}  ,~\left\{  X_{2}\right\}  ,~\left\{  X_{3}\right\}
,~\left\{  X_{1}+\gamma X_{2}\right\}  ~,~\left\{  X_{1}+\delta X_{\Psi
}\right\}  ,~\\
&  \left\{  X_{2}+\delta X_{\Psi}\right\}  ~,~\left\{  X_{3}+\delta X_{\Psi
}\right\}  ~\text{and }\left\{  X_{1}+\gamma X_{2}+\delta X_{\Psi}\right\}  ~
\end{align*}
Hence, we shall determine seven invariant solutions for the WdW equations
(\ref{WD.02}) which are not related through adjoint transformation%

\begin{table}[tbp] \centering
\caption{Commutators of the admitted Lie point symmetries for the WdW equation \ref{WD.02}.}%
\begin{tabular}
[c]{ccccc}\hline\hline
$\left[  ~,~\right]  $ & $\mathbf{X}_{1}$ & $\mathbf{X}_{2}$ & $\mathbf{X}%
_{3}$ & $\mathbf{X}_{4}$\\
$\mathbf{X}_{1}$ & $0$ & $0$ & $X_{1}$ & $0$\\
$\mathbf{X}_{2}$ & $0$ & $0$ & $-X_{2}$ & $0$\\
$\mathbf{X}_{3}$ & $-X_{1}$ & $X_{2}$ & $0$ & $0$\\
$\mathbf{X}_{4}$ & $0$ & $0$ & $0$ & $0$\\\hline\hline
\end{tabular}
\label{tabl1}%
\end{table}%
%

\begin{table}[tbp] \centering
\caption{Adjoint representation of the admitted Lie point symmetries for the WdW equation \ref{WD.02}.}%
\begin{tabular}
[c]{ccccc}\hline\hline
$Ad\left(  e^{\left(  \varepsilon\mathbf{X}_{i}\right)  }\right)
\mathbf{X}_{j}$ & $\mathbf{X}_{1}$ & $\mathbf{X}_{2}$ & $\mathbf{X}_{3}$ &
$\mathbf{X}_{4}$\\
$\mathbf{X}_{1}$ & $X_{1}$ & $X_{2}$ & $-\varepsilon X_{1}+X_{3}$ & $X_{4}$\\
$\mathbf{X}_{2}$ & $X_{1}$ & $X_{2}$ & $\varepsilon X_{2}+X_{3}$ & $X_{4}$\\
$\mathbf{X}_{3}$ & $e^{\varepsilon}X_{1}$ & $e^{-\varepsilon}X_{2}$ & $X_{3}$
& $X_{4}$\\
$\mathbf{X}_{4}$ & $X_{1}$ & $X_{2}$ & $X_{3}$ & $X_{4}$\\\hline\hline
\end{tabular}
\label{tabl2}%
\end{table}

By using $\left\{  X_{1}\right\}  $ and $X_{2}$ we infer that $\Psi\left(
z,w\right)  =0,$ which is a trivial solution. On the other hand by
using$~\left\{  X_{3}\right\}  $ we find
\begin{equation}
\Psi_{3}\left(  z,w\right)  =\Psi_{3\left(  1\right)  }^{0}I_{0}\left(
\frac{w^{2}\sqrt{z}}{3}\right)  +\Psi_{3\left(  2\right)  }^{0}K_{0}\left(
\frac{w^{2}\sqrt{z}}{3}\right)  ,
\end{equation}
where $I_{0}\left(  x\right)  ,~K_{0}\left(  x\right)  $ are the modified
Bessel functions and $\Psi_{3\left(  1\right)  }^{0},~\Psi_{3\left(  2\right)
}^{0}$ are constants.

In addition, from the symmetry vector $\left\{  X_{1}+\gamma X_{2}\right\}  $
we calculate the travel-wave like wavefunction%
\begin{equation}
\Psi_{12}\left(  z,w\right)  =\Psi_{12\left(  1\right)  }^{0}\exp\left(
i\frac{w^{4}-4\gamma z}{12\sqrt{\gamma}}\right)  +\Psi_{12\left(  1\right)
}^{0}\exp\left(  -i\frac{w^{4}-4\gamma z}{12\sqrt{\gamma}}\right)  .
\end{equation}

In a similar way, the rest of the similarity solutions are determined to be
\begin{equation}
\left\{  X_{1}+\delta X_{\Psi}\right\}  :\bar{\Psi}_{1}\left(  z,w\right)
=\Psi_{1}^{0}\exp\left(  \delta z+\frac{w^{4}}{36\delta}\right)
\end{equation}%
\begin{equation}
\left\{  X_{2}+\delta X_{\Psi}\right\}  :\bar{\Psi}_{2}\left(  z,w\right)
=\Psi_{2}^{0}\exp\left(  \frac{z}{9\delta}+\frac{w^{4}\delta}{4}\right)
\end{equation}%
\begin{equation}
\left\{  X_{3}+\delta X_{\Psi}\right\}  :\bar{\Psi}_{3}\left(  z,w\right)
=\left(  w^{-2\delta}z^{\frac{\delta}{2}}\right)  \left(  \bar{\Psi}_{3\left(
1\right)  }^{0}I_{\delta}\left(  \frac{w^{2}\sqrt{z}}{3}\right)  +\bar{\Psi
}_{3\left(  2\right)  }^{0}K_{\delta}\left(  \frac{w^{2}\sqrt{z}}{3}\right)
\right)
\end{equation}
and
\begin{align}
\left\{  X_{1}+\gamma X_{2}+\delta X_{\Psi}\right\}   &  :\bar{\Psi}%
_{4}\left(  z,w\right)  =\Psi_{12\left(  1\right)  }^{0}\exp\left(
\frac{\left(  3\delta+i\sqrt{4\gamma-9\delta^{2}}\right)  \left(
w^{4}-4\gamma z\right)  }{24\gamma}+\delta z\right)  +\nonumber\\
&  +\Psi_{12\left(  2\right)  }^{0}\exp\left(  \frac{\left(  3\delta
-i\sqrt{4\gamma-9\delta^{2}}\right)  \left(  w^{4}-4\gamma z\right)
}{24\gamma}+\delta z\right)  .
\end{align}

We observe that solutions $\bar{\Psi}_{1}\left(  z,w\right)  ,~\Psi_{2}\left(
z,w\right)  $ \thinspace and $\Psi_{12}\left(  z,w\right)  ~$are equivalent,
hence we have found in total five independent similarity solutions. Because
the WdW equation is linear the generic similarity solution by point
transformations is written as%
\begin{equation}
\Psi\left(  z,w\right)  =%
{\displaystyle\sum}
\bar{\alpha}_{1}\bar{\Psi}_{1}\left(  z,w\right)  +%
{\displaystyle\sum}
\alpha_{3}\Psi_{3}\left(  z,w\right)  +%
{\displaystyle\sum}
\bar{a}_{3}\bar{\Psi}_{3}\left(  z,w\right)  +%
{\displaystyle\sum}
\bar{a}_{4}\bar{\Psi}_{4}\left(  z,w\right)  ,
\end{equation}
where the sum is on all the free parameters of the solutions. Recall that no
boundary conditions have been applied to constrain the similarity solutions.
The boundary conditions in quantum cosmology is still an open problem. \ 

However, for the classical system and specifically from (\ref{WD.01}) the
Hamilton-Jacobi equation follows $\frac{\partial S}{\partial z}\frac{\partial
S}{\partial w}-\frac{w^{3}}{9}=0~$with the constraint equation $\frac{\partial
S}{\partial z}=S_{0}$, which is nothing else than the conservation law
$\dot{p}_{z}=0$. Consequently, the generic solution of the Hamilton-Jacobi
equation is
\begin{equation}
S\left(  z,w\right)  =S_{0}z+\frac{w^{4}}{36S_{0}}%
\end{equation}
which is nothing else than the exponent function of the similarity solution
$\bar{\Psi}_{1}\left(  z,w\right)  $. \ Therefore, we can infer that solution
$\bar{\Psi}_{1}\left(  z,w\right)  $ is the one which recovers the classical
solution where parameter $\delta$ is related with the conservation law
$p_{z}=S_{0}$. \ 

\begin{figure}[ptb]
\includegraphics[height=8cm]{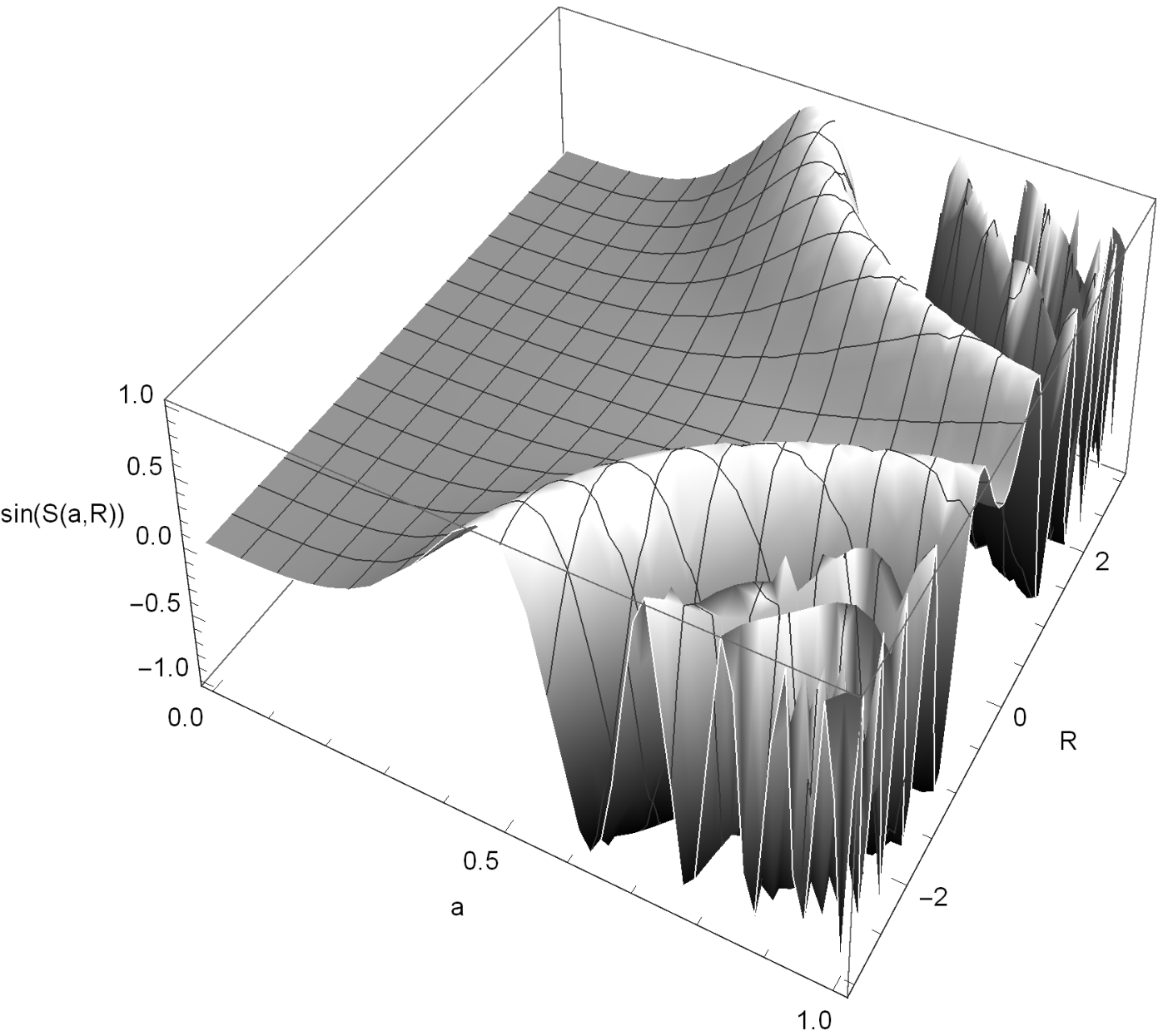}\centering\caption{Qualitative
evolution of the wavefunction $\operatorname{Im}\left(  \bar{\Psi}_{1}\left(
a,R\right)  \right)  $ for $\delta=\frac{i}{10},$ that is,~ $\operatorname{Im}%
\left(  \bar{\Psi}_{1}\left(  a,R\right)  \right)  \sim\sin\left(  S\left(
a,R\right)  \right)  $ with $S\left(  a,R\right)  $ be the solution of the
Hamilton-Jacobi equation.}%
\label{fig01}%
\end{figure}

In addition, we observe that the solution of the Hamilton-Jacobi equation is
included in~solution $\bar{\Psi}_{4}\left(  z,w\right)  $, but not in the rest
of the solutions, namely $\Psi_{3}\left(  z,w\right)  $ and $\bar{\Psi}%
_{3}\left(  z,w\right)  $. \ In Fig. \ref{fig01} we give the qualitative
evolution of the wavefunction $\operatorname{Im}\left(  \bar{\Psi}_{1}\left(
a,R\right)  \right)  $ for $\delta=\frac{i}{10},$ that is,~ $\operatorname{Im}%
\left(  \bar{\Psi}_{1}\left(  a,R\right)  \right)  \sim\sin\left(  S\left(
a,R\right)  \right)  $ where $S\left(  a,R\right)  $ is the solution of the
Hamilton-Jacobi equation. 

\subsection{Case B: Power law model $R^{\frac{7}{8}}$}

For the power-law model $f\left(  R\right)  =R^{\frac{7}{8}}$ we prefer to
work on the new \ coordinates $\left\{  \rho,\sigma\right\}  $
\begin{equation}
a=\left(  \frac{21}{4}\right)  ^{-\frac{1}{3}}\sqrt{\rho e^{\sigma}%
}\;\;,\;R=\frac{e^{12\sigma}}{\rho^{4}},
\end{equation}
where the point-like Lagrangian takes the simple form%
\begin{equation}
L\left(  \rho,\dot{\rho},\sigma,\dot{\sigma}\right)  =\frac{1}{2}\dot{\rho
}^{2}-\frac{1}{2}\rho^{2}\dot{\sigma}^{2}+V_{0}\frac{e^{12\sigma}}{\rho^{2}}.
\end{equation}
The latter Lagrangian describes the two-dimensional Ermakov-Pinney system
without the oscillatory term, while constant $V_{0}$ has the value
$V_{0}=-\frac{1}{42}$.

The Hamiltonian constraint is
\begin{equation}
H=\frac{1}{2}p_{\rho}^{2}-\frac{1}{2\rho^{2}}\left(  p_{\sigma}^{2}%
-2V_{0}e^{12\sigma}\right)  \equiv0, \label{le.00}%
\end{equation}
where%
\begin{equation}
\Phi_{0}=\left(  p_{\sigma}^{2}-2V_{0}e^{12\sigma}\right)  , \label{le.01}%
\end{equation}
is the Ermakov-Pinney invariant, also known as Lewis invariant.%

\begin{table}[tbp] \centering
\caption{Commutators of the admitted Lie point symmetries for the WdW equation \ref{WD.06}.}%
\begin{tabular}
[c]{ccccc}\hline\hline
$\left[  ~,~\right]  $ & $\mathbf{Y}_{1}$ & $\mathbf{Y}_{2}$ & $\mathbf{Y}%
_{3}$ & $\mathbf{Y}_{4}$\\
$\mathbf{Y}_{1}$ & $0$ & $-6Y_{2}$ & $6Y_{3}$ & $0$\\
$\mathbf{Y}_{2}$ & $6Y_{2}$ & $0$ & $0$ & $0$\\
$\mathbf{Y}_{3}$ & $-6Y_{3}$ & $0$ & $0$ & $0$\\
$\mathbf{Y}_{4}$ & $0$ & $0$ & $0$ & $0$\\\hline\hline
\end{tabular}
\label{tabl3}%
\end{table}%
%

\begin{table}[tbp] \centering
\caption{Adjoint representation of the admitted Lie point symmetries for the WdW equation \ref{WD.06}.}%
\begin{tabular}
[c]{ccccc}\hline\hline
$Ad\left(  e^{\left(  \varepsilon\mathbf{Y}_{i}\right)  }\right)
\mathbf{Y}_{j}$ & $\mathbf{Y}_{1}$ & $\mathbf{Y}_{2}$ & $\mathbf{Y}_{3}$ &
$\mathbf{Y}_{4}$\\
$\mathbf{Y}_{1}$ & $Y_{1}$ & $e^{6\varepsilon}Y_{2}$ & $e^{-6\varepsilon}%
Y_{3}$ & $Y_{4}$\\
$\mathbf{Y}_{2}$ & $Y_{1}-6\varepsilon Y_{2}$ & $Y_{2}$ & $Y_{3}$ & $Y_{4}$\\
$\mathbf{Y}_{3}$ & $Y_{1}+6\varepsilon Y_{3}$ & $Y_{2}$ & $Y_{3}$ & $Y_{4}$\\
$\mathbf{Y}_{4}$ & $Y_{1}$ & $Y_{2}$ & $Y_{3}$ & $Y_{4}$\\\hline\hline
\end{tabular}
\label{tabl4}%
\end{table}%

As far as the WdW equation (\ref{wdw.01}) is concerned it is calculated to be
\begin{equation}
\Psi_{\rho\rho}-\frac{1}{\rho^{2}}\Psi_{\sigma\sigma}+\frac{1}{\rho}\Psi
_{\rho}-2V_{0}\frac{e^{12\sigma}}{\rho^{2}}\Psi=0. \label{WD.06}%
\end{equation}
The later partial differential equation is invariant under the one-parameter
point transformations with generators the vector fields%
\begin{align*}
~Y_{1}  &  =\rho\partial_{\rho}~,~Y_{2}=\rho^{-5}e^{-6\sigma}\partial_{\rho
}+\rho^{-6}e^{-6\sigma}\partial_{\sigma}~,\\
Y_{3}  &  =\rho^{7}e^{-6\sigma}\partial_{\rho}-\rho^{6}e^{-6\sigma}%
\partial_{\sigma}~,~Y_{\Psi}=\Psi\partial_{\Psi}.
\end{align*}
where in the canonical forms are%
\begin{align*}
~\hat{Y}_{1}  &  =\rho\Psi_{\rho}\partial_{\Psi}~,~\hat{Y}_{2}=\rho
^{-5}e^{-6\sigma}\left(  \Psi_{\rho}+\Psi_{\sigma}\right)  \partial_{\Psi
}+\rho^{-6}e^{-6\sigma}\partial_{\sigma}~,\\
\hat{Y}_{3}  &  =\rho^{6}e^{-6\sigma}\left(  \rho\Psi_{\rho}-\Psi_{\sigma
}\right)  \partial_{\Psi}~,~\hat{Y}_{\Psi}=\Psi\partial_{\Psi}.
\end{align*}

In tables \ref{tabl3} and \ref{tabl4} we present the commutators and the
adjoint representation of the admitted point symmetries. From table
\ref{tabl4} we find that the one-dimensional optimal system to be%
\begin{align*}
&  \left\{  Y_{1}\right\}  ,~\left\{  Y_{2}\right\}  ~,~\left\{
Y_{3}\right\}  ~,~\left\{  Y_{2}-\gamma Y_{3}\right\}  ~,~\left\{
Y_{1}+\delta Y_{4}\right\}  ~,\\
&  \left\{  Y_{2}+\delta Y_{4}\right\}  ~,~\left\{  Y_{3}+\delta
Y_{4}\right\}  ~,~\left\{  Y_{2}-\gamma Y_{3}+\delta Y_{4}\right\}  .
\end{align*}
From the one-dimensional algebras $\left\{  Y_{2}\right\}  $ and $\left\{
Y_{3}\right\}  $ we find the trivial solutions $\Psi\left(  \rho
,\sigma\right)  =0$. From the other one-dimensional algebras it follows%
\begin{equation}
\left\{  Y_{1}\right\}  :\Psi_{1}\left(  \rho,\sigma\right)  =\Psi_{1\left(
1\right)  }^{0}I_{0}\left(  \frac{\sqrt{21}}{126}e^{6\sigma}\right)
+\Psi_{1\left(  2\right)  }^{0}K_{0}\left(  \frac{\sqrt{21}}{126}e^{6\sigma
}\right)  \label{ss1}%
\end{equation}%
\begin{equation}
\left\{  Y_{2}-\gamma Y_{3}\right\}  :\Psi_{23}\left(  \rho,\sigma\right)
=\Psi_{23\left(  1\right)  }^{0}\sin\left(  \frac{\sqrt{21}}{252\sqrt{\gamma}%
}e^{6\zeta}\right)  +\Psi_{23\left(  2\right)  }^{0}\sin\left(  \frac
{\sqrt{21}}{252\sqrt{\gamma}}e^{6\zeta}\right)  ~,~\zeta=y+\frac{1}{6}%
\ln\left(  \frac{\left(  \gamma\rho^{12}-1\right)  }{\rho^{6}}\right)
\end{equation}%
\begin{equation}
\left\{  Y_{1}+\delta Y_{4}\right\}  :\bar{\Psi}_{1}\left(  \rho
,\sigma\right)  =\rho^{\delta}\left(  \Psi_{1\left(  1\right)  }^{0}%
I_{\frac{\delta}{6}}\left(  \frac{\sqrt{21}}{126}e^{6\sigma}\right)
+\Psi_{1\left(  2\right)  }^{0}K_{\frac{\delta}{6}}\left(  \frac{\sqrt{21}%
}{126}e^{6\sigma}\right)  \right)
\end{equation}%
\begin{equation}
\left\{  Y_{2}+\delta Y_{4}\right\}  :\bar{\Psi}_{2}\left(  \rho
,\sigma\right)  =\bar{\Psi}_{2}^{0}\exp\left(  \frac{\delta}{12}\rho^{6}%
e^{6y}+\frac{1}{252\delta}\rho^{-6}e^{6y}\right)  ,
\end{equation}%
\begin{equation}
\left\{  Y_{3}+\delta Y_{4}\right\}  :\bar{\Psi}_{3}\left(  \rho
,\sigma\right)  =\bar{\Psi}_{2}^{0}\exp\left(  -\frac{\delta}{12}\rho
^{-6}e^{6y}-\frac{1}{252\delta}\rho^{6}e^{6y}\right)  ,
\end{equation}
while from $\left\{  Y_{3}+\delta Y_{4}\right\}  $ we get the solution
of~$\Psi_{23}\left(  \rho,\sigma\right)  $ multiplied by \ the function
$\exp\left(  \frac{\delta}{12\gamma}\rho^{-6}e^{6y}\right)  \,$.

However as we discussed in the previous section for the Ermakov-Pinney system
the Lewis invariant (\ref{le.01}) can be used to construct the
Lie-B\"{a}cklund operator%
\begin{equation}
\Psi_{\sigma\sigma}+2V_{0}e^{12}\Psi=\left(  c_{J}\right)  ^{2}\Psi
\end{equation}
By using the later constraint we find the wavefunction%
\begin{equation}
\Psi_{LB}\left(  \rho,\sigma\right)  =\left(  a_{1}\rho^{c_{J}}+a_{2}%
\rho^{-c_{J}}\right)  \left(  \left[  b_{1}J_{\frac{c_{J}}{6}}\left(  \frac
{1}{6}\sqrt{2V_{0}}e^{6\sigma}\right)  +b_{2}Y_{\frac{c_{J}}{6}}\left(
\frac{1}{6}\sqrt{2V_{0}}e^{6\sigma}\right)  \right]  \right)  \label{s01}%
\end{equation}
where $a_{1},a_{2},b_{1}$ and $b_{2}$ are integration constants and
$J_{i}\left(  x\right)  ,~Y_{i}\left(  x\right)  $ are the Bessel functions.
We observe that solutions $\Psi_{1}\left(  \rho,\sigma\right)  $ and
$\bar{\Psi}_{1}\left(  \rho,\sigma\right)  $ are included in the latter
generic solution. In total we have found four different solutions, hence, the
generic wavefunction is expressed as%
\begin{equation}
\Psi\left(  \rho,\sigma\right)  =%
{\displaystyle\sum}
\bar{\alpha}_{1}\Psi_{LB}+%
{\displaystyle\sum}
\alpha_{23}\Psi_{23}\left(  \rho,\sigma\right)  +%
{\displaystyle\sum}
\bar{a}_{2}\bar{\Psi}_{2}+%
{\displaystyle\sum}
\bar{\alpha}_{23}\bar{\Psi}_{23}\left(  \rho,\sigma\right)  .
\end{equation}

In order to relate any quantum solution with the classical universe, we should
solve the Hamilton-Jacobi equation (\ref{le.00}) with the use of the
constraint (\ref{le.01}) where $p_{\rho}=\frac{\partial S}{\partial\rho}$ and
$p_{\sigma}=\frac{\partial S}{\partial\sigma}$. We find that%
\begin{equation}
S\left(  \rho,\sigma\right)  =\sqrt{\Phi_{0}}\ln\rho+\frac{1}{6}\sqrt
{2V_{0}e^{12\sigma}+\Phi_{0}}+\frac{\Phi_{0}}{6}\arctan h\left(  \frac
{\sqrt{2V_{0}e^{12\sigma}+\Phi_{0}}}{\Phi_{0}}\right)  ,~\Phi_{0}\neq0,
\end{equation}
or%
\begin{equation}
S\left(  \rho,\sigma\right)  =-\frac{\sqrt{2V_{0}}}{6}e^{6\sigma},~\Phi_{0}=0.
\end{equation}

We observe that there is not any direct relation between the similarity
solutions for the WdW equation and the Hamilton-Jacobi for the classical
system. However, if we focus on the limits of the Bessel functions we shall
see that the classical limit is recovered.

\begin{figure}[ptb]
\includegraphics[height=8cm]{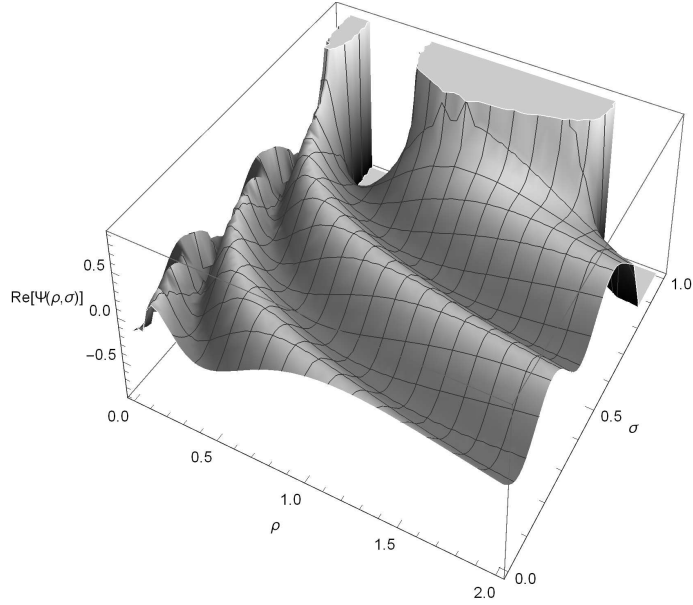}\centering\caption{Qualitative
evolution of the real part of the wavefunction (\ref{s01}) $\operatorname{Re}%
\left(  \Psi_{LB}\left(  a,R\right)  \right)  $ for $c_{J}=3i.$}%
\label{fig2}%
\end{figure}

Consider the similarity solution $\Psi_{LB}\left(  \rho,\sigma\right)
,~$\ with $c_{J}=i\sqrt{\Phi_{0}}$ and $a_{2}$, then in the limit$~e^{6\sigma
}\rightarrow+\infty$ it follows that
\begin{equation}
\Psi_{LB}\left(  \rho,\sigma\right)  \simeq e^{-3\sigma}e^{i\left(  \sqrt
{\Phi_{0}}\ln\rho+\sqrt{2V_{0}}e^{6\sigma}\right)  .}%
\end{equation}
which is actually the imaginary exponent of the wavefunction correspond to the
limit of $S\left(  \rho,\sigma\right)  $ as $e^{6\sigma}\rightarrow+\infty$.
Hence, we can see that the classical limit is recovered. The qualitative
evolution of the similarity solution $\Psi_{LB}\left(  \rho,\sigma\right)  $
is presented in Fig. \ref{fig2} for $c_{J}=3i$.

\subsection{Case C: Model $\left(  R-2\Lambda\right)  ^{\frac{3}{2}}$}

For the third model of analysis, namely model C with $f(R)=(R-2\Lambda)^{3/2}%
$, we work on the coordinates $\left\{  z,w\right\}  $ similar to that of
model A, that is%
\[
a=\left(  \frac{9}{2}\right)  ^{-\frac{1}{3}}\sqrt{z}\;\;\;\;R=2\Lambda
+\frac{w^{2}}{z}%
\]
the point-like Lagrangian is written in the canonical form as
\begin{equation}
L\left(  z,\dot{z},w,\dot{w}\right)  =\dot{z}\dot{w}+\frac{1}{9}w^{3}%
+\omega^{2}zw, \label{c.01}%
\end{equation}
where parameter $\omega$ is defined as $\omega=\sqrt{2\Lambda/3}$. The term
with coefficient $\omega^{2}$ in the latter point-like Lagrangian it is an
oscillator term, that can be easily seen if someone writes the latter
Lagrangian in diagonal coordinates.

Hence, from (\ref{c.01}) it follows that the Hamiltonian constraint is%
\begin{equation}
H=p_{z}p_{w}-\frac{1}{9}w^{3}-\omega^{2}zw\equiv0,
\end{equation}
while the field equations are%
\begin{equation}
\dot{z}=p_{w}~~~~~\dot{w}=p_{z},
\end{equation}%
\begin{equation}
\dot{p}_{z}=\omega^{2}w~~~~~\dot{p}_{w}=\frac{1}{3}w^{2}+\omega^{2}z.
\end{equation}
From the latter system we construct the quadratic conservation law
$I_{0}=p_{z}^{2}-\omega^{2}w^{2}\,.$

The solution of the Hamilton-Jacobi equation by using the quadratic
conservation law $I_{0}$ is found to be%
\begin{equation}
S\left(  z,w\right)  =\frac{\sqrt{I_{0}+\omega^{2}w^{2}}}{\omega^{4}}\left(
\omega^{2}w^{2}+27\omega^{4}z-2I_{0}\right)  \label{c.01a}%
\end{equation}

The WdW equation for this specific model is written in the coordinates
$\left\{  z,w\right\}  $ as%
\begin{equation}
\Psi_{zw}-\left(  \frac{1}{9}w^{3}+\omega^{2}zw\right)  \Psi=0. \label{c.02}%
\end{equation}
The linear partial differential equation (\ref{c.02}) is invariant under the
point transformations with infinitesimal generators the vector fields%
\[
Z_{1}=2\partial_{z}-\frac{9}{w}\omega^{2}\partial_{w}~,~Z_{\Psi}=\Psi
\partial_{\Psi}.
\]
The one-dimensional optimal system consists of by the vector fields $\left\{
Z_{1}\right\}  ,~\left\{  Z_{1}+\delta Z_{\Psi}\right\}  $. In canonical form
the vector field $Z_{1}$ is written as $\hat{Z}_{1}=\left(  2\Psi_{z}-\frac
{9}{w}\omega^{2}\Psi_{w}\right)  \partial_{\Psi}$.

From the point transformation $\left\{  Z_{1}+\delta Z_{\Psi}\right\}  $ the
similarity solution follows%
\begin{equation}
\Psi_{1}\left(  z,w\right)  =\exp\left(  \frac{\delta}{4}z-\frac{\delta
}{36\omega^{2}}w^{2}\right)  \left(  \Psi_{1\left(  1\right)  }^{0}Ai\left(
\zeta\right)  +\Psi_{1\left(  2\right)  }^{0}Bi\left(  \zeta\right)  \right)
~, \label{c.03}%
\end{equation}
where $Ai\left(  \zeta\right)  ,~Bi\left(  \zeta\right)  $ are the Airy
functions and $\zeta=-\frac{6^{\frac{2}{3}}}{288\omega^{\frac{8}{3}}}\left(
1+\sqrt{3}i\right)  \left(  \delta^{2}+72\omega^{4}z+8\omega^{2}w^{2}\right)
$. It is not a surprise that the wavefunction is expressed by the Airy
functions. Recall that the Airy functions solve the Schr\"{o}dinger equation
for a particle confined by a triangular well \cite{air}.

However, by using the differential operator generated by the quadratic
conservation law $I_{0}$, that is,%
\begin{equation}
\hat{I}_{0}\Psi\equiv\Psi_{zz}-\omega^{2}w^{2}\Psi+c_{J}\Psi
\end{equation}
we find the similarity solution%
\begin{equation}
\Psi_{LB}\left(  z,w\right)  =\Psi_{LB\left(  1\right)  }^{0}\sin\left(
\xi\right)  +\Psi_{LB\left(  2\right)  }^{0}\cos\left(  \xi\right)
\end{equation}
where parameter $\xi$ is defined as%
\begin{equation}
\xi=\frac{\sqrt{c_{J}-\omega^{2}w^{2}}}{\omega^{4}}\left(  \omega^{2}%
w^{2}+27\omega^{4}z+2c_{J}\right)  . \label{c.04}%
\end{equation}
Consequently, we can see that $\xi\left(  w,z\right)  $ is nothing else than
the solution of the Hamilton-Jacobi for the classical system (\ref{c.01a}).
Therefore we observe that the classical solution is recovered by the
wavefunction $\Psi_{LB}\left(  z,w\right)  $. \ 

\begin{figure}[ptb]
\includegraphics[height=8cm]{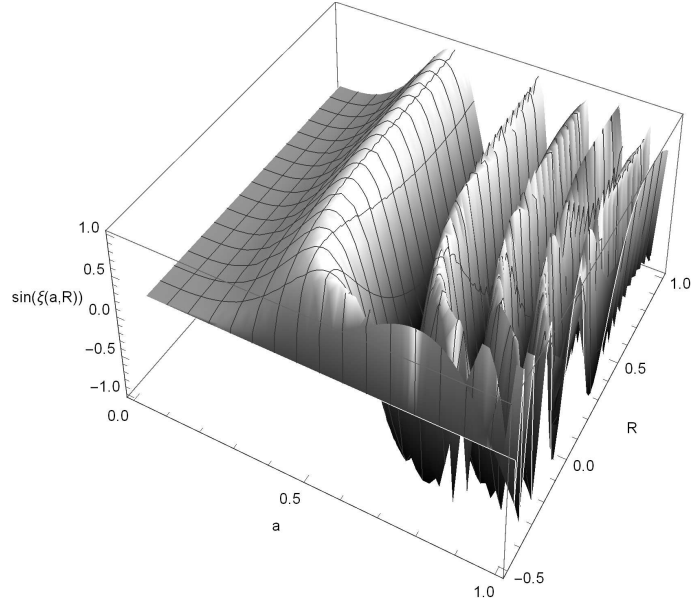}\centering\caption{Qualitative
evolution of the wavefunction $\Psi_{LB}\left(  a,R\right)  $ for negative
value of $\Lambda$ and specifically for $\Lambda=-\frac{1}{5}$ .}%
\label{fig3}%
\end{figure}

In Fig. \ref{fig3} the qualitative evolution of $\Psi_{LB}\left(  a,R\right)
$ is presented for negative value of $\Lambda$ and~$\Psi_{LB\left(  2\right)
}^{0}=0$ 

\subsection{Case D: Model $\left(  R-2\Lambda\right)  ^{\frac{7}{8}}$}

Model $f\left(  R\right)  =\left(  R-2\Lambda\right)  ^{\frac{7}{8}}$,
describes the Ermakov-Pinney system with a nonzero oscillator term. Indeed in
the coordinates $\left\{  \rho,\sigma\right\}  $the point-like Lagrangian for
the field equations is written as%
\begin{equation}
L\left(  \rho,\dot{\rho},\sigma,\dot{\sigma}\right)  =\frac{1}{2}\dot{\rho
}^{2}-\frac{1}{2}\rho^{2}\dot{\sigma}^{2}+V_{0}\frac{m}{4}\rho^{2}+V_{0}%
\frac{e^{12\sigma}}{\rho^{2}}%
\end{equation}
where $\bar{m}=-28\Lambda~,~V_{0}=-\frac{1}{42}$,~and
\begin{equation}
a=\left(  \frac{21}{4}\right)  ^{-\frac{1}{3}}\sqrt{\rho e^{\sigma}%
}\;\;\;\;R=2\Lambda+\frac{e^{12\sigma}}{\rho^{4}}.
\end{equation}

In the new coordinates, the Hamiltonian constraint is written
\begin{equation}
H\equiv\frac{1}{2}\dot{\rho}^{2}-\frac{1}{2}\rho^{2}\dot{\sigma}^{2}%
-V_{0}\frac{m}{4}\rho^{2}-V_{0}\frac{e^{12\sigma}}{\rho^{2}}=0
\end{equation}
while the field equations becomes%
\begin{equation}
\dot{\rho}=p_{\rho},~\dot{\sigma}=\frac{p_{\sigma}}{\rho^{2}}~,~\dot
{p}_{\sigma}=\frac{12V_{0}}{\rho^{2}}e^{12\sigma}%
\end{equation}%
\begin{equation}
\dot{p}_{\rho}=-\frac{1}{\rho^{3}}p_{\sigma}^{2}+\frac{V_{0}m}{2}\rho
-\frac{2V_{0}}{\rho^{3}}e^{12\sigma}.
\end{equation}
Finally, the field equations admit the Lewis invariant which is written as
\begin{equation}
\Phi=\dot{\sigma}^{2}+V_{0}e^{12\sigma}.
\end{equation}

The WdW equation (\ref{wdw.01}) is written as follows%
\begin{equation}
\Psi_{\rho\rho}-\frac{1}{\rho^{2}}\Psi_{\sigma\sigma}+\frac{1}{^{\rho}}%
\Psi_{\rho}-2\left(  V_{0}\frac{m}{4}\rho^{2}+V_{0}\frac{e^{12\sigma}}%
{\rho^{2}}\right)  \Psi=0, \label{d.01}%
\end{equation}
and has no other point symmetries except the trivial ones. However, as we
discussed in Section \ref{sec3} from the Lewis-invariant we construct the
differential operator%
\begin{equation}
\hat{\Phi}\Psi\equiv\Psi_{\sigma\sigma}+V_{0}e^{12\sigma}\Psi-\Phi_{0}\Psi.
\label{d.02}%
\end{equation}

\begin{figure}[ptb]
\includegraphics[height=8cm]{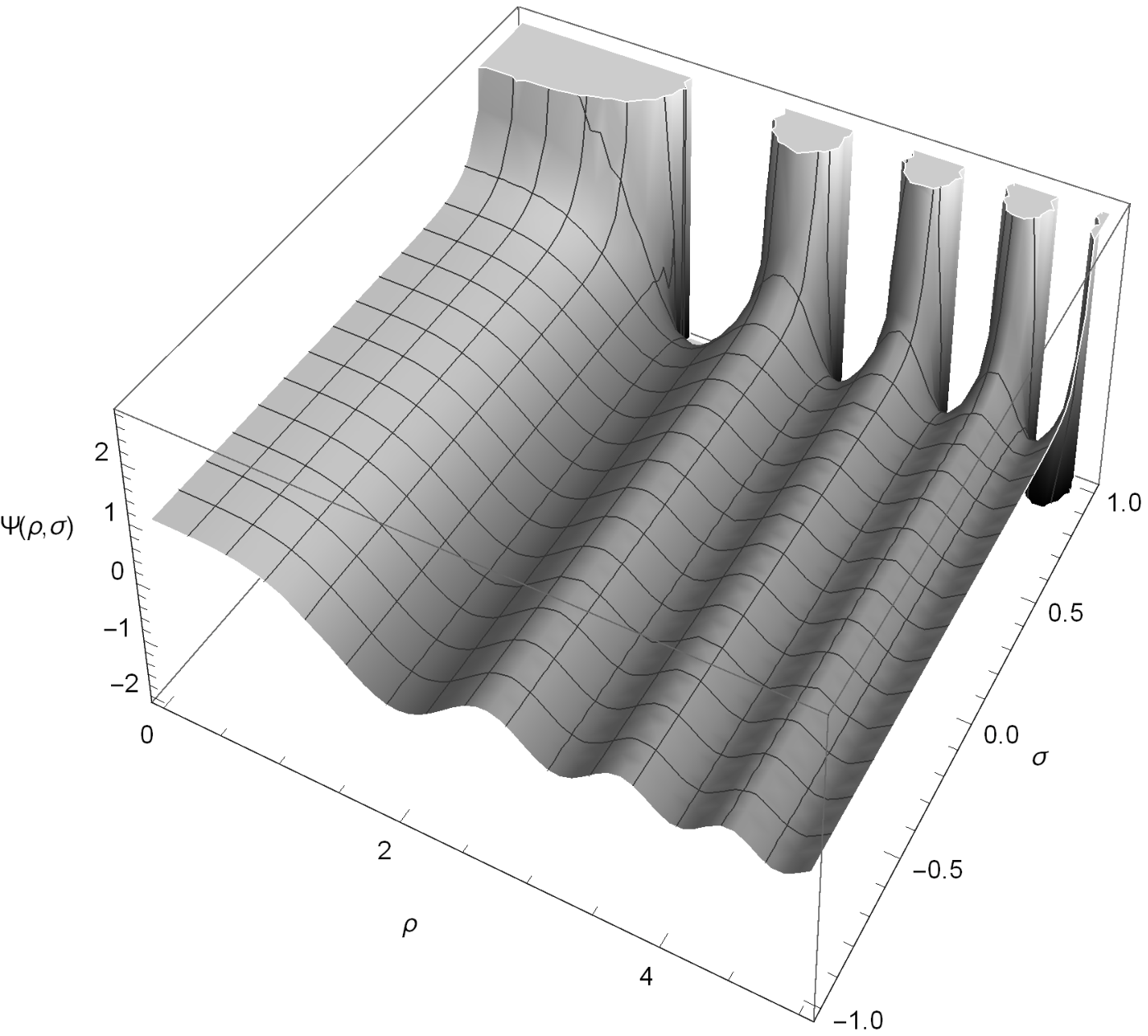}\centering\caption{Qualitative
evolution of the wavefuction $\Psi_{LB}\left(  \rho,\sigma\right)  $ for
$\Psi_{LB\left(  2\right)  }^{0}=\Psi_{LB\left(  3\right)  }^{0}%
=\Psi_{LB\left(  4\right)  }^{0}=0$ and for $\Lambda<0.$}%
\label{fig4}%
\end{figure}

Hence from (\ref{d.01}) and (\ref{d.02}), with $\hat{\Phi}\Psi=0$ we find the
similarity solution%
\begin{align}
\Psi_{LB}\left(  \rho,\sigma\right)   &  =\left(  \Psi_{LB\left(  1\right)
}^{0}I_{\frac{\sqrt{\Phi_{0}}}{6}}\left(  \frac{\sqrt{42}}{252}e^{6\sigma
}\right)  +\Psi_{LB\left(  2\right)  }^{0}K_{\frac{\sqrt{\Phi_{0}}}{6}}\left(
\frac{\sqrt{42}}{252}e^{6\sigma}\right)  \right)  J_{\frac{\sqrt{\Phi_{0}}}%
{2}}\left(  \frac{\sqrt{-6\Lambda}}{12}\rho^{2}\right)  +\nonumber\\
&  +\left(  \Psi_{LB\left(  3\right)  }^{0}I_{\frac{\sqrt{\Phi_{0}}}{6}%
}\left(  \frac{\sqrt{42}}{252}e^{6\sigma}\right)  +\Psi_{LB\left(  4\right)
}^{0}K_{\frac{\sqrt{\Phi_{0}}}{6}}\left(  \frac{\sqrt{42}}{252}e^{6\sigma
}\right)  \right)  Y_{\frac{\sqrt{\Phi_{0}}}{2}}\left(  \frac{\sqrt{-6\Lambda
}}{12}\rho^{2}\right)  \label{d.03}%
\end{align}
where,~$I_{a}\left(  x\right)  ,$ $J_{a}\left(  x\right)  $,$~K_{a}\left(
x\right)  $ and $Y_{a}\left(  x\right)  ~$are the Bessel functions. \ We
observe that in order the wavefunction to be total periodic, then $\Lambda<0$.
As far as concerns the classical limit, in a similar approach with Model B,
that is recovered in the limit where $e^{6\sigma}\rightarrow+\infty$. The
qualitative evolution of $\Psi_{LB}\left(  \rho,\sigma\right)  $ for
$\Psi_{LB\left(  2\right)  }^{0}=\Psi_{LB\left(  3\right)  }^{0}%
=\Psi_{LB\left(  4\right)  }^{0}=0$ and for $\Lambda<0$ is presented in Fig.
\ref{fig4}.

\section{Conclusions}

\label{sec5}

In this work we focused on the determination of similarity solutions for the
WdW equation in quantum cosmology and more specifically in $f\left(  R\right)
$-gravity in a spatially flat FLRW universe. The WdW equation is a linear
equation of Klein-Gordon class which by definition is conformal invariant. For
the cosmological of our consideration the WdW equation provides the solution
of the wavefunction $\Psi$ in terms of the two indepedent variables of the
theory, the scale factor $a\left(  t\right)  $ and the Ricciscalar $R\left(
t\right)  $.

We recall, that $f\left(  R\right)  $ is a fourth-order theory and the
Ricciscalar $R\left(  t\right)  $ has been added as a Lagrangian multiplier in
order to attribute the higher-order derivatives, such that the $f\left(
R\right)  -$gravity to be of second-order but with more degrees of freedom.
Because of the latter property the theory is dynamical equivalent with
scalar-tensor theories while a point-like Lagrangian description is possible,
which is necessary for our approach on the problem.

For the function form of $f\left(  R\right)  $ which specifies the theory, we
considered four models which were found before and are integrable by
one-parameter point transformations. Two of the models are power-law while the
other two models belong to the family of $\Lambda_{bc}$CDM. For these specific
models we write the WdW equation and we determine the infinitesimal generators
of the one-parameter transformations where the WdW equations are invariant. We
use the infinitesimal generators to define Lie-B\"{a}cklund operators which
are used as constraint equations to solve the WdW equation. These solutions
are called similarity solutions.

A novel observation for the solutions that were found by that approach is that
in the classical limit, that is, in the WKB approximation, the solution of the
Hamilton-Jacobi equation for the classical system is recovered, consequently
the classical limit is recovered. We can say that the similarity solutions
which provide the classical limit are preferred. Indeed there are not initial
and boundary conditions to constrain the solutions of the WdW equation,
however by the requirement the similarity solution to provide the classical
limit we can construct a family of boundary conditions. Because a similarity
solution is invariant under the infinitesimal transformations which have been
applied for the determination, the boundary conditions should be also
invariant under the same infinitesimal transformations \cite{bd1,bd2}.

The similarity solutions can be used to define probability, or calculate the
quantum potential of Bohmiam mechanics. However such applications is not the
scope of the present work and such analysis will be published elsewhere.

\begin{acknowledgments}
The author acknowledges Marianthi Paliathanasi for continuous support.
\end{acknowledgments}

\end{document}